# The story of ALICE: Building the dedicated heavy ion detector at LHC[1]

C. Fabjan[a] and J. Schukraft
CERN, 1211 Geneva 23, Switzerland

[a] now at Inst. of High Energy Physics, Austrian Academy of Sciences and Vienna University of Technology

This article documents the main design choices and the close to 20 years of preparation, detector R&D, construction and installation of ALICE, the dedicated heavy ion experiment at the CERN LHC accelerator.

**Introduction**

ALICE, which stands for A Large Ion Collider Experiment, is very different in both design and purpose from the other experiments at the LHC. Its main aim is the study of head-on collisions between heavy nuclei, at first mainly lead on lead at the top energy of the LHC. In these reactions, the LHC's enormous energy – collisions of Pb nuclei are 100 times more energetic than those of protons – heats up the matter in the collision zone to a temperature which is 100,000 times higher than the temperature in the core of our sun. Nuclei and nucleons melt into their elementary constituents, quarks and gluons, to form for a brief instant the primordial matter which filled the universe until a few microseconds after the Big Bang.
The hot reaction zone expands at almost the speed of light, and in the process cools, breaks up and condenses back into a plethora of ordinary, composite matter particles.
The ALICE detector has to measure as many as possible of the escaping particles, totalling up to several tens of thousands in each of these 'little bangs', and record their number, type, mass, energy and direction, in order to infer from the debris the existence and properties of matter under the extreme conditions created during the instance of the collision.
ALICE will also take data with protons in order to compare them with the heavy ion results, both to look for telltale chances between the two types of beams and to characterise the global event structure of proton reactions with its very different and complementary set of detectors.

**History and Challenges**

The physics program of high-energy heavy-ion collisions and the search for the 'Quark-Gluon Plasma'(QGP), the primordial matter of the Universe, started in 1986 at the CERN SPS accelerator and, simultaneously, at the Brookhaven AGS in the US. The first set of detectors, many of them put together from equipment used in previous generations of experiments, could actually only use rather light ion beams (from Oxygen to Silicon) and what today would be considered a rather modest energy at these fixed target machines (5 - 20 GeV centre-of-mass energy). Already the following year, in 1987, during a workshop to choose CERN's next accelerator project amongst different contenders, the possibility of using both heavy ions as well as protons was mentioned for the machine which was to become the LHC. In 1990, when in the US the Relativistic Heavy Ion Collider (RHIC) - dedicated to heavy ion physics - was approved and a call for experiments was issued, the European community faced the decision to either participate at RHIC or focus its resources on the LHC. The schedules for RHIC and LHC were, at the time, quite comparable and therefore a sequential exploitation of both machines seemed impossible. After a series of workshops and discussions, which looked both at the physics potential of these machines and at different detector concepts, the decision was made, correct as it would seem with hindsight, to participate from Europe at a modest scale at RHIC and to start in parallel a dedicated design and R&D effort for a large general purpose heavy ion detector at LHC. This left the community with a busy schedule and thinly stretched resources: ongoing data analysis of the light ion program , constructing a new generation of experiments for the heavy ion program starting at CERN with Pb beams in 1994, designing and building detectors for RHIC (in operation since 2000), and R&D for an ambitious experiment at LHC.  All this went on in parallel and involved many of the same actors and groups, but it did pay off handsomely in a rich program and fast progress.

---
[1] This article has been prepared for the experiment chapter of the book *'The Large Hadron Collider: A marvel technology'*, EPFL-Press Lausanne, Switzerland, 2009 (Editor: L. Evans).

Designing a dedicated heavy ion experiment in the early 90's for use at the LHC some 15 years later posed some daunting challenges: In a field still in its infancy, it required extrapolating the conditions to be expected by a factor of 300 in energy and a factor of 7 in beam mass. The detector therefore had to be both 'general purpose' – able to measure most signals of potential interest, even if their relevance may only become apparent later – and flexible, allowing additions and modifications along the way as new avenues of investigation would open up. In both respects ALICE did quite well, as it included a number of observables in its initial menu whose importance only became clear after results appeared from RHIC, and various major detection systems where added over time, from the muon spectrometer in 1995, the transition radiation detector in 1999,  to a large jet calorimeter added as recently as 2007.

Other challenges relate to the experimental conditions expected for nucleus-nucleus collisions at the LHC. The most difficult one to meet is the extreme number of particles produced in every single event, which could be up to three orders of magnitude larger than in typical proton-proton interactions at the same energy and a factor two to five still above the highest multiplicities measured at RHIC. The tracking of these particles was therefore made particularly safe and robust by using mostly three-dimensional hit information with many points along each track (up to 150) in a moderate magnetic field (too strong a field would both mix up the particles and exclude the lowest energy ones from being observed).
In addition, a large dynamic range is required for momentum measurement, spanning more than three orders of magnitude from tens of MeV to well over 100 GeV. This is achieved with a combination of very low material thickness (to reduce scattering of low momentum particles) and a large tracking lever arm of up to 3.5 m  (resolution improves at high momentum with the square of the measurement length), thus achieving good resolution at both high and low momentum with modest field.
And finally, Particle Identification (PID) over much of this momentum range is essential, as many phenomena depend critically on either particle mass or particle type. ALICE therefore employs essentially all known PID techniques in a single experiment, as discussed in some detail in the following sections.

The ALICE design evolved from the Expression of Interest (1992) via a Letter of Intent (1993) to the Technical Proposal (1996) and was officially approved in 1997.  The first ten years were spent on design and an extensive R&D effort. Like for all other LHC experiments, it became clear from the outset that also the challenges of heavy ion physics at LHC could not be really met (nor paid for) with existing technology. Significant advances, and in some cases a technological break-through, would be required to built on the ground what physicists had dreamed up on paper for their experiments. The initially very broad and later more focused, well organised and well supported R&D effort, which was sustained over most of the 1990's, has lead to many evolutionary and some revolutionary advances in detectors, electronics and computing.

# ALICE Detector Overview

ALICE is usually referred to as one of the smaller detectors, but the meaning of 'small' is very relative in the context of LHC: The detector stands 16 meters tall, is 16 m wide and 26 m long, and weights in at approximately 10,000 tons. It has been designed and built over almost two decades by a collaboration which currently includes over 1000 scientists and engineers from more than 100 Institutes in some 30 different countries. The experiment consists of 17 different detection systems, each with its own specific technology choice and design constraints.

A schematic view of ALICE is shown in Fig.1. It consists of a central part, which measures hadrons, electrons, and photons, and a forward single arm spectrometer that focuses on muon detection. The central 'barrel' part covers the direction perpendicular to the beam from 45° to 135° and is located inside a huge solenoid magnet, which was built in the 1980's for an experiment at CERN's LEP accelerator. As a warm resistive magnet, the maximum field at the nominal power of 4 MW reaches 0.5 T. The central barrel contains a set of tracking detectors, which record the momentum of the charged particles by measuring their curved path inside the magnetic field. These particles are then identified according to mass and particle type by a set of particle identification detectors, followed by two types of electromagnetic calorimeters for photon and jet measurements. The forward muon arm (2°-9°) consists of a complex arrangement of absorbers, a large dipole magnet, and fourteen planes of tracking and triggering chambers.

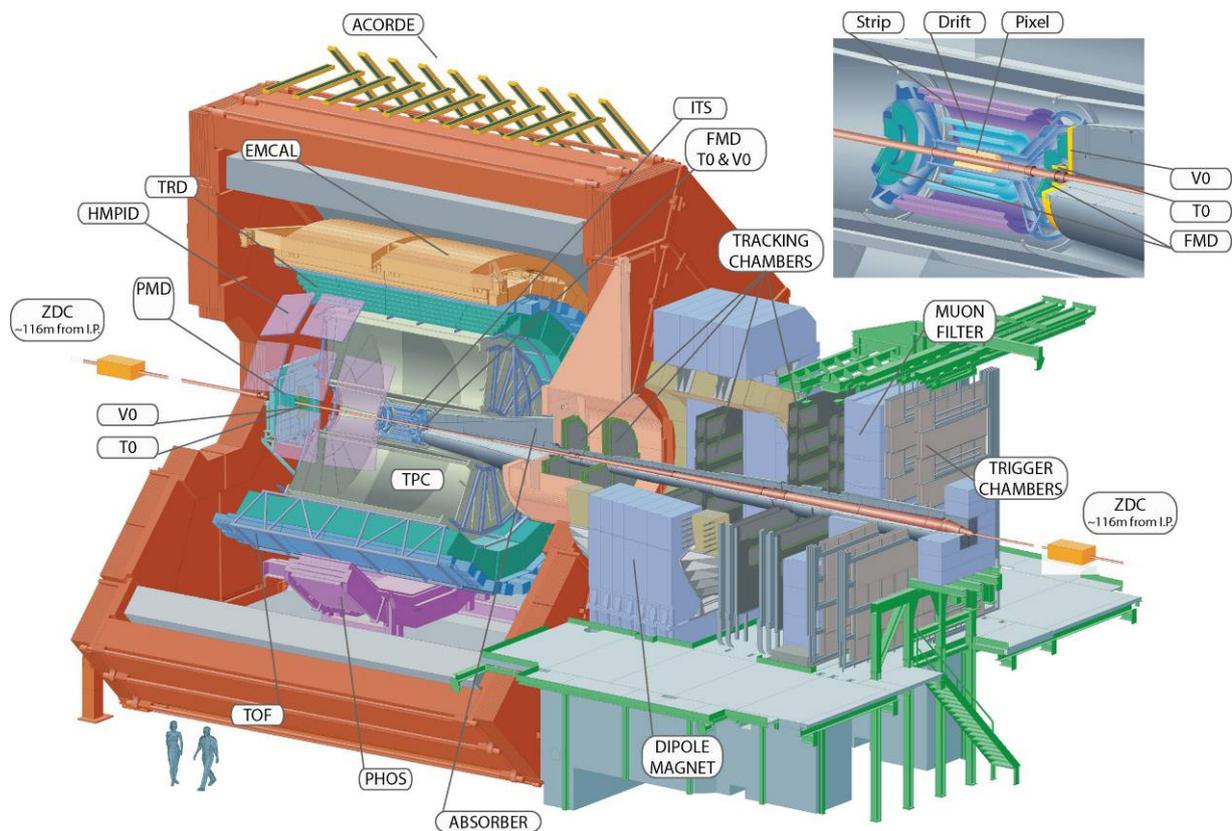

Fig 1: Computer model of the ALICE detector. The different sub-detectors are labelled with their acronym, which are not always explained but listed both in the figure and the text to help locating the various components of ALICE.

**Observing the 'Primordial Matter' relics**

The ultimate goal of ALICE is to track the remnants of the primordial matter, a task that requires the combination of three very challenging detection capabilities:
- *Reconstruction* of all the tracks of tens of thousands of particles;
- *Measurement* of the momentum of these particles from very low (100 MeV/c) to very high ($\geq$ 100 GeV/c) values;
- *Identification* of most particles through their specific interaction with different detectors;
- *Observation* of the decay vertices, a fraction of a millimeter away from the collision, of the telltale heavy *charm* and *bottom quarks*.

Figure 2 shows a typical example of how the primordial matter might reveal itself in the Alice Tracker. The key to meeting this challenge is the combination of state-of-the-art tracking with specially developed low-mass silicon detectors and a very low–mass time projection chamber optimized for this high multiplicity environment, followed by a suite of detectors specialised on identifying the particles. The different technologies are described in the following sections, starting with the detectors closest to the point of (inter)action.

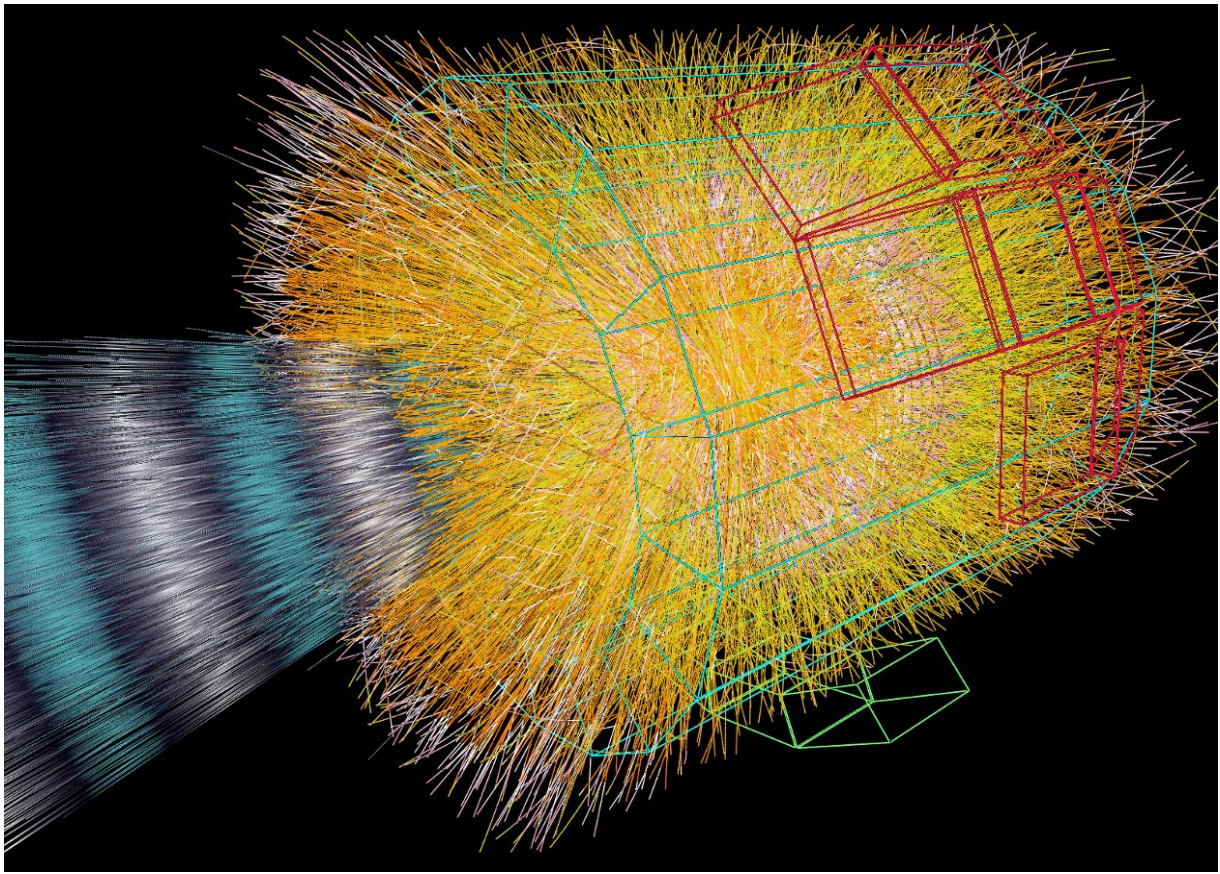

Fig 2: A simulated high-multiplicity event detected in the Alice tracking detectors

**Tracking detectors:**

Tracking in the central barrel is divided into the Inner Tracking System (ITS), a six-layer, silicon vertex detector, and the Time-Projection Chamber (TPC). The ITS locates the primary vertex (where the collision occurs) and secondary vertices (where some of the unstable heavy particles decay after a flight distance of some hundreds of micrometers) with a precision on the order of a few tens of micrometers. Because of the high particle density, the innermost four layers need to be high resolution devices, i.e. silicon *pixel detectors* and silicon *drift*

*detectors*, which record both x and y coordinates for each passing particle. The outer layers are equipped with double-sided silicon *micro-strip detectors*. The total area covered with silicon detectors reaches 7 m$^2$ and includes almost 13 million individual electronic channels.

The Silicon Pixel Detector (SPD)

The LHC experiments pioneered this novel detector technology. This is a 'checker-board' detector with tiny detection elements, typically 0.05 mm by 0.5 mm, resulting in a huge number (10 to 100 million) of detection channels. These detectors offer the best particle tracking capability of all presently existing methods. A host of problems had to be solved: connectivity between the pixel detector and the pixel readout chip ('bump bonding'); dedicated electronics microchips to amplify and digitize the signal and to serialize the data; cooling of detectors with the constraint of low mass; low mass support structures.

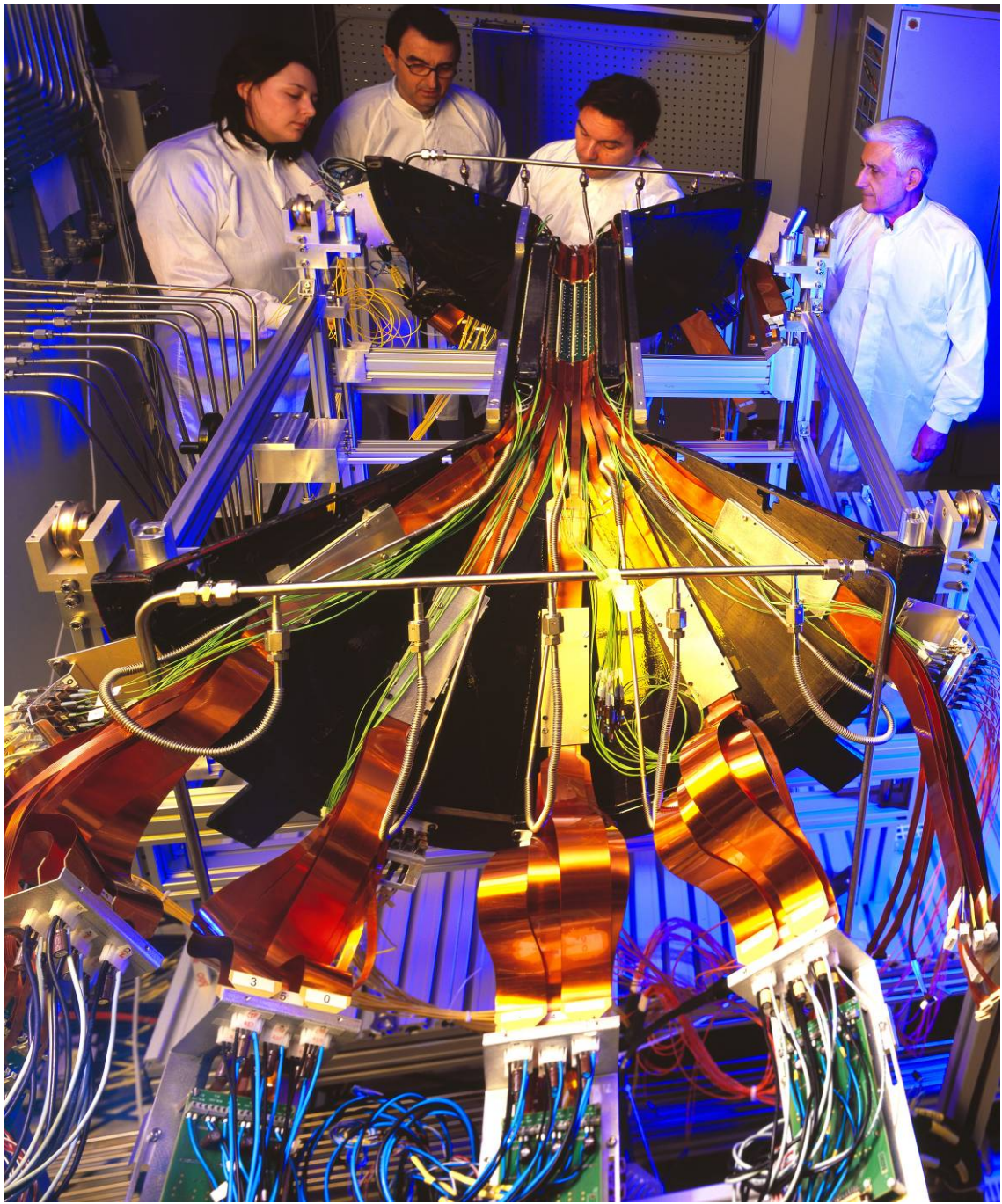

Fig 3. Partial view of one half of the SPD during installation. The service connections dwarf the actual pixel detector (© A. Saba).

The ALICE SPD is assembled from ten interlocking carbon-fiber structures, carrying two layers of pixels, cooling pipes and readout connections. In the obsession to minimize materials the pixel detector wafers were manufactured with a thickness of only 0.2 mm; the electronic readout chips were mechanically thinned – after the bump bonding – from 0.3 mm to 0.15 mm (!); novel readout buses were developed, replacing the technology standard of copper readout with aluminium tracks, again aiming to minimize the material of the detector. This effort has been rewarded by creating a pixel detector representing a world record in lowest total mass.

The ALICE SPD has a further unique feature, not found in the other LHC pixel detectors: it provides a trigger on charged particles within less than 900 ns, in time for the next-level trigger decision. It will be used as an *interaction trigger*, including the possibility to select very unusual multiplicity configurations. The partially installed SPD can be seen in Fig 3, predominantly showing the conical fan-out of services to avoid occulting detectors near the beam pipe.

The Silicon Drift Detector (SDD)

The particle density behind the SPD is still so high that the subsequent tracker must also provide unambiguous two-dimensional space points with pixel-like space resolution. Apart from the pixel technology only the silicon drift chamber concept provides this feature and it was selected by the ALICE collaboration as the best cost-performance option. This technology was never used on a large scale, although the original idea is more than 20 years old. The R&D had to address three main areas: the fabrication of suitable Si-drift detectors, the readout electronics and connectivity, and stability of operation. The relatively large size (88 x 73 mm2) of the 0.3 mm thick detector units contributed significantly to the production difficulties. A total of 260 such modules are installed in the completed detector. The electrical and cooling connectivity posed again one of the major technological hurdles; low-mass cables were 'de rigeur', implying the development of aluminium signal tracks on Kapton. This work, carried out in the Ukraine, was almost 'torpedoed' by over-zealous customs officials, who decided that the readout buses were highly sensitive military material. As drift detectors are very sensitive to temperature variations, they will be both thermo-stabilized to within a fraction of a degree and (in a belt and suspenders approach) monitored by numerous electronic structures distributed densely over the surface of the detectors.

Time Projection Chamber (TPC)

The need for efficient and robust tracking has led to the choice of a TPC as the main tracking detector. In spite of its drawbacks concerning slow recording speed and huge data volume, only such a tracking device providing highly redundant information can guarantee reliable performance with tens of thousands of charged particles within the geometrical *acceptance*.
A conceptual view of the TPC is shown in Fig.4. Charged particles leave ionizing tracks in the huge gas filled cylinder, which is separated into two halves by the central electrode. The liberated electrons slowly drift in a strong electric field of 100 kV towards the two ends of the TPC, where they are amplified and recorded in wire chambers (only one of these chambers is shown in the figure on the left side). By measuring many space points along each track - the arrival point in the chamber gives two coordinates whereas the distance of the track from the endplate is inferred from drift time - a TPC is ideally suited to disentangle the dense web of particle tracks in heavy ion reactions.

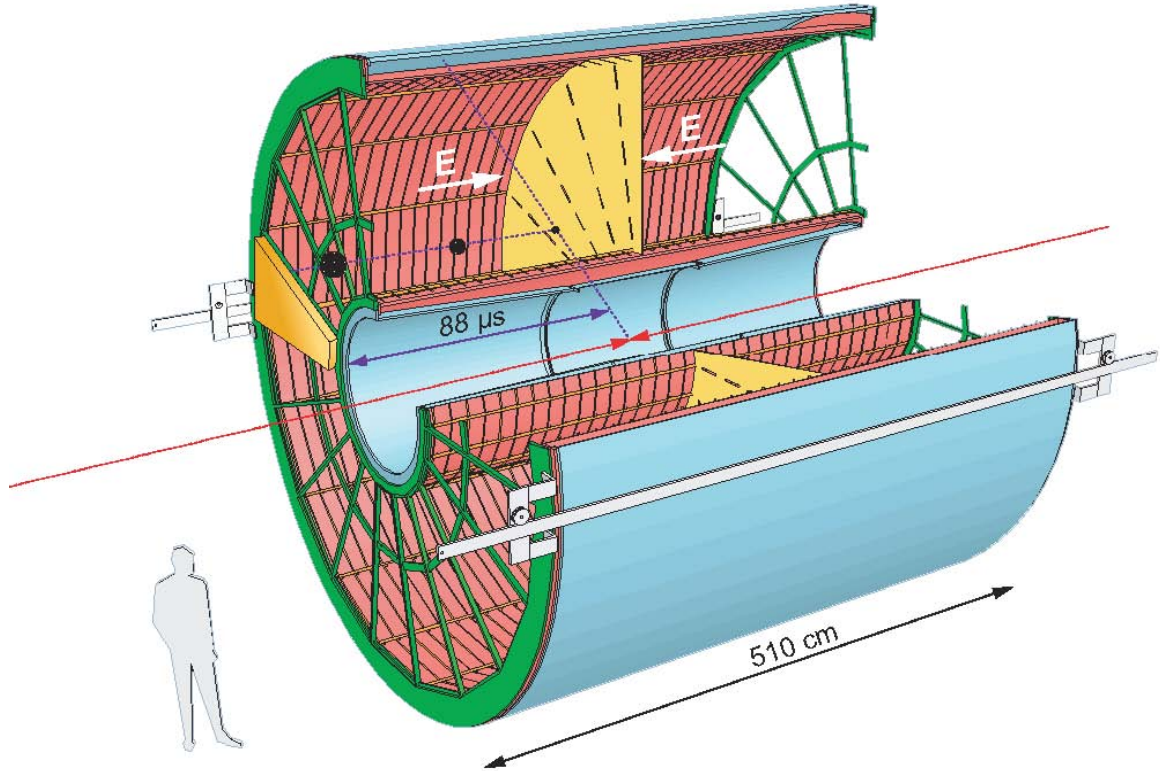

Fig. 4: Conceptual view of the TPC, showing the dimensions and components.

This world's biggest TPC may be considered as the 'workhorse' of ALICE, providing by itself already a tremendous physics reach. Again, ALICE aimed for an exceptionally low-mass detector by making the appropriate choice of materials and gas, combined with tight dimensional and environmental controls.

The 5.6 m diameter and 5.4 m long field cage is built from two Carbon-fiber/honeycomb composite cylinders, materials normally used for space-applications. Dimensional tolerances which are critical for the performance, such as the uniformity of the drift field, were kept at the $10^{-4}$ level.

The central drift electrode has been built with a planarity and parallelism to the readout chambers of better than 0.2 mm, see the artistic view of the inner 90 m$^3$, Fig. 5. The second component to an ultralow-mass TPC is the choice of chamber gas: the conventional Argon gas was rejected as unacceptably massive in favor of lighter Neon. The final TPC is the lightest ever constructed, representing approximately the same amount of material as the ITS.

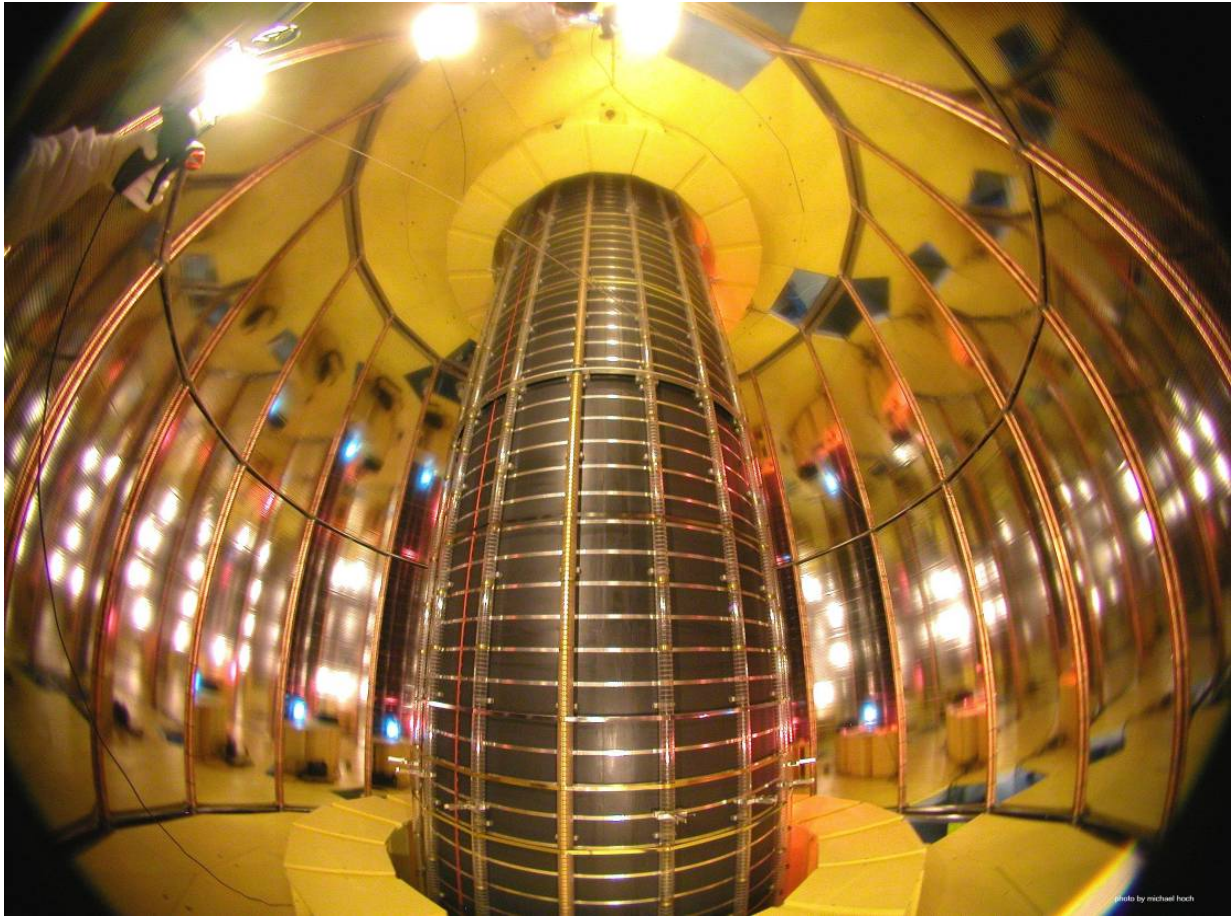

Fig. 5: Inner tracking volume of the TPC. The layered structure provides a constant electric field of 400 V/cm between the central electrode and the readout chambers (© CERN).

The second innovation is the readout electronics: a preamplifier/ signal shaper, operating at the fundamental thermal limit of noise, is followed by a specially developed readout chip, the ALICE Tpc Read Out (ALTRO) chip. It processes digitally the signals for optimized performance at high collision rates. The electronics is miniaturized to a level that it is fully mounted on the end plates of the TPC, connected merely by 260 optical fibers to the Data Acquisition System (Fig 6). This readout is rapidly becoming the reference for gaseous detectors.

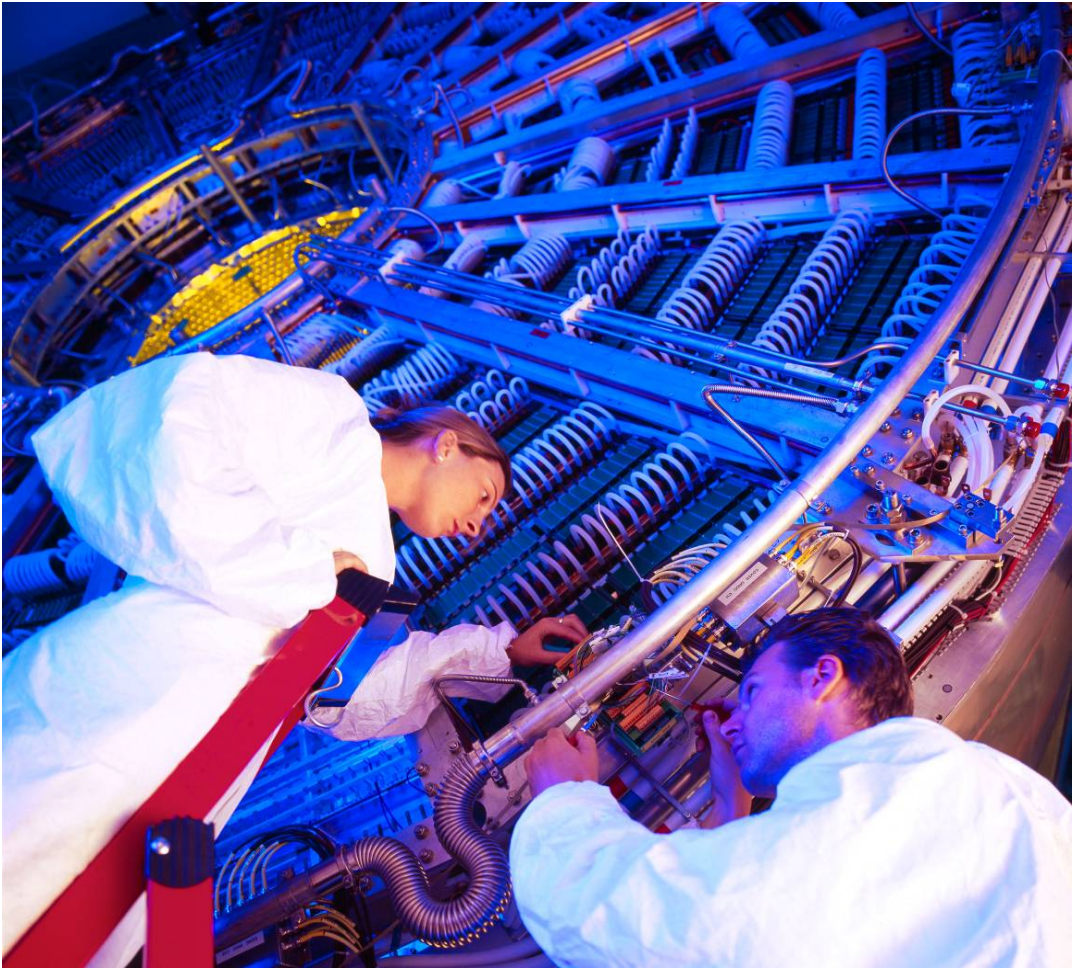

Fig. 6: Installation of the TPC readout electronics (© A. Saba).

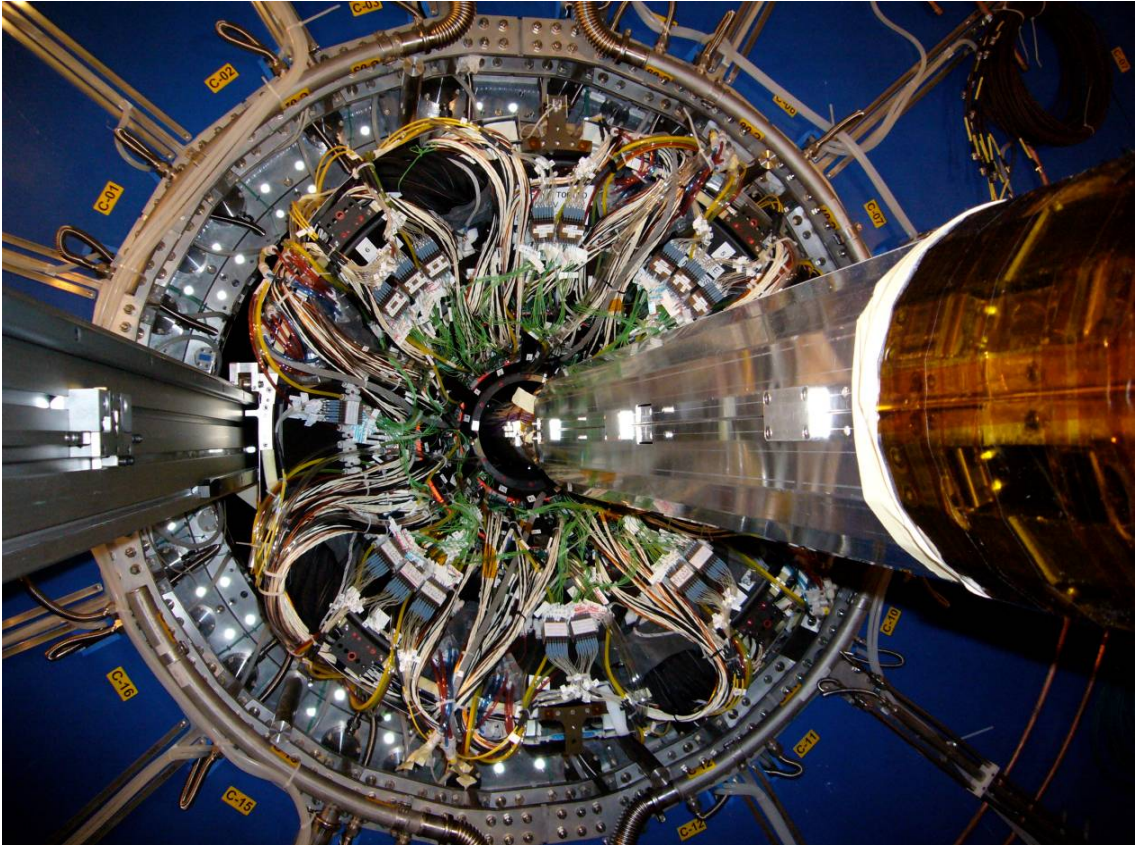

Fig. 7: View of the TPC and ITS during the installation (© CERN).

The performance advantage of Neon as the drift gas is not for free: environmental conditions (temperature and pressure) have to be controlled tightly, given the strong dependence of drift velocity on pressure and temperature. The needed $10^{-4}$ control on drift velocity requires a temperature stability of ~ $0.1^{\circ}$ C over the volume of 90 $m^3$. Needless to say that extreme thermal stabilization measures were employed to reach this goal. Fig 7 shows the TPC during the installation.

**Particle identification detectors**

Particle identification over a large part of the phase space and for many different particles is an important design feature of ALICE with several detector systems dedicated to PID:
- the Time-of-Flight (TOF) array measures the flight time of particles from the collision point out to the detector; together with the momentum this time determines the mass.
- The High Momentum Particle Identifier Detector (HMPID) uses Cherenkov radiation to measure particle velocities very close to the speed of light.
- The Transition Radiation Detector (TRD) will identify electrons above 1 GeV to study production rates of heavy quarks (charm and beauty mesons).

In addition to these dedicates detectors, the energy loss in the tracking detectors ITS and TPC is also used for PID; the former only good at low momentum (< 1 GeV/c), the latter out to very high momenta (> 60 GeV/c).

<u>The Time-of-Flight System (TOF)</u>

The Time-of-Flight system has the function to distinguish between pions, kaons and protons, up to several GeV/c, needing a quantum jump in technology. The PID performance required a ~ $5 \times 10^{-11}$s (50 ps) time resolution, in the presence of the extreme particle multiplicities. Such timing performance had been reached in small systems with scintillators, which however are not practical and too expensive for 150 $m^2$ of detector area and more than 150,000 detection channels. Initially, several different methods were explored without much success. Finally a breakthrough was reached with one venerable technique – the 'Resistive Plate Chamber' – through understanding of the detection process, insight and perseverance. The detector is made from 10 detection gaps of 0.2 mm each, delivering a staggering performance of 50 ps, a factor 50 improvement compared to conventional RPC's. The high channel number implied the development of novel microelectronics for the amplification and time measurement with a precision of 25 ps. A new TOF standard was born and is starting to revolutionize particle identification (Fig. 8).

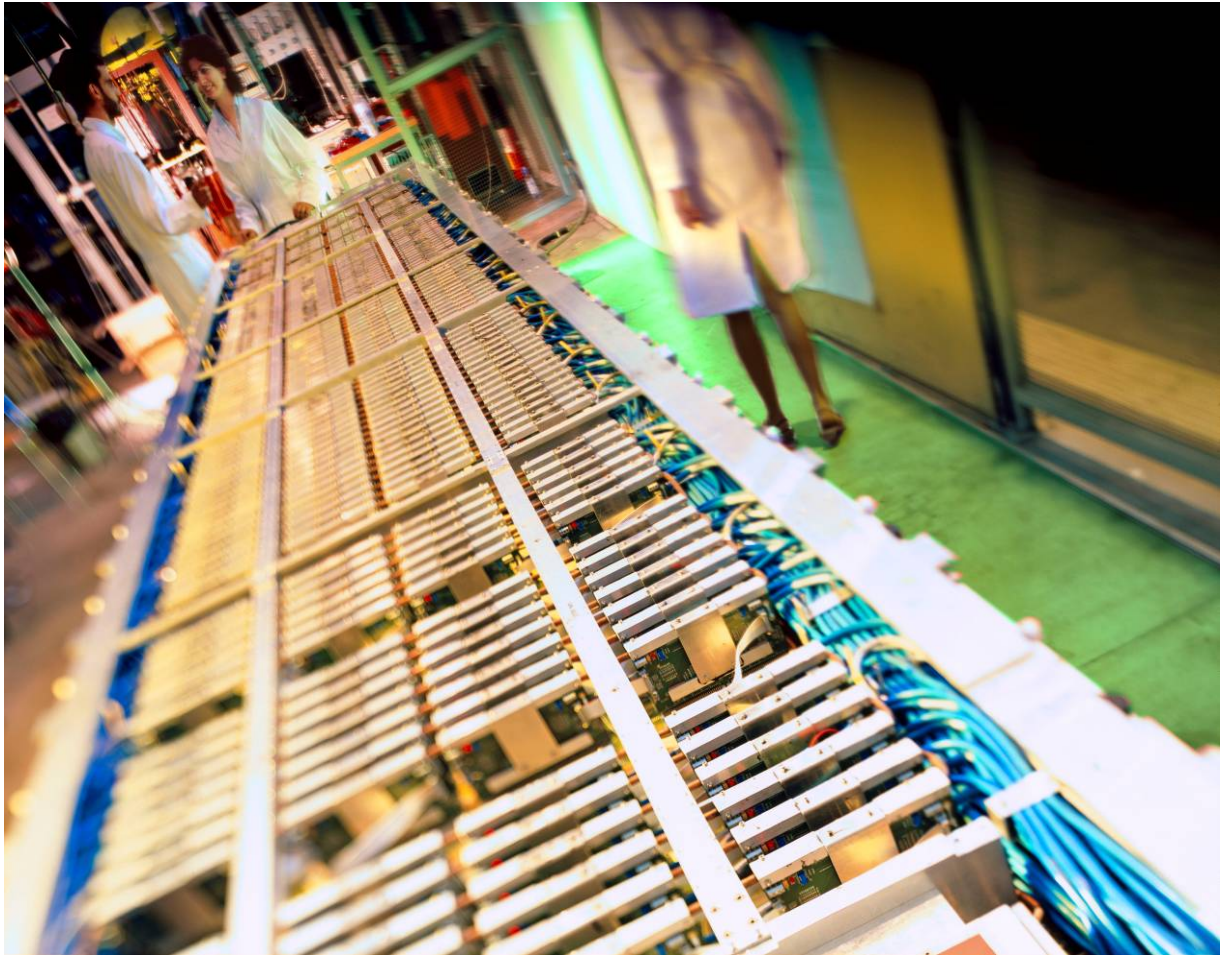

Fig 8: View of one of 18 TOF modules during construction (© A. Saba).

The High Momentum Particle Identifier (HMPID).

The evolution of another method, to which many groups have contributed over the past 30 years, was pushed to new performance standards. It is the 'Cherenkov Detector': Under certain conditions, particles traversing a medium (e.g. a gas or liquid) excite the medium to emit 'Cherenkov light', named after the discoverer of this phenomenon. The light is emitted under a characteristic angle, determined by the velocity of the particle. Detecting the very faint light and its direction, measures the velocity which allows deducing the mass and hence the type of particle. This is the task of the HMPID. The concept is as beautifully simple as it is as fiendishly hard to turn it into a practical detector. The difficulty resides in detecting the very few (20 to 30) emitted photons with detectors which register also the position of the photons with mm – accuracy. The HMPID collaboration solved this problem by perfecting the production of large-area photocathodes, based on Caesium iodide (CsI) films, evaporated onto gas detector electrode surfaces of square meter size. A photon impinging on the CsI surface will produce with high probability (~25 %) an electron which can be detected. This development provided the basic ingredient for one of the world's largest Imaging Cherenkov Detectors, ever built for an accelerator experiment, Fig. 9.

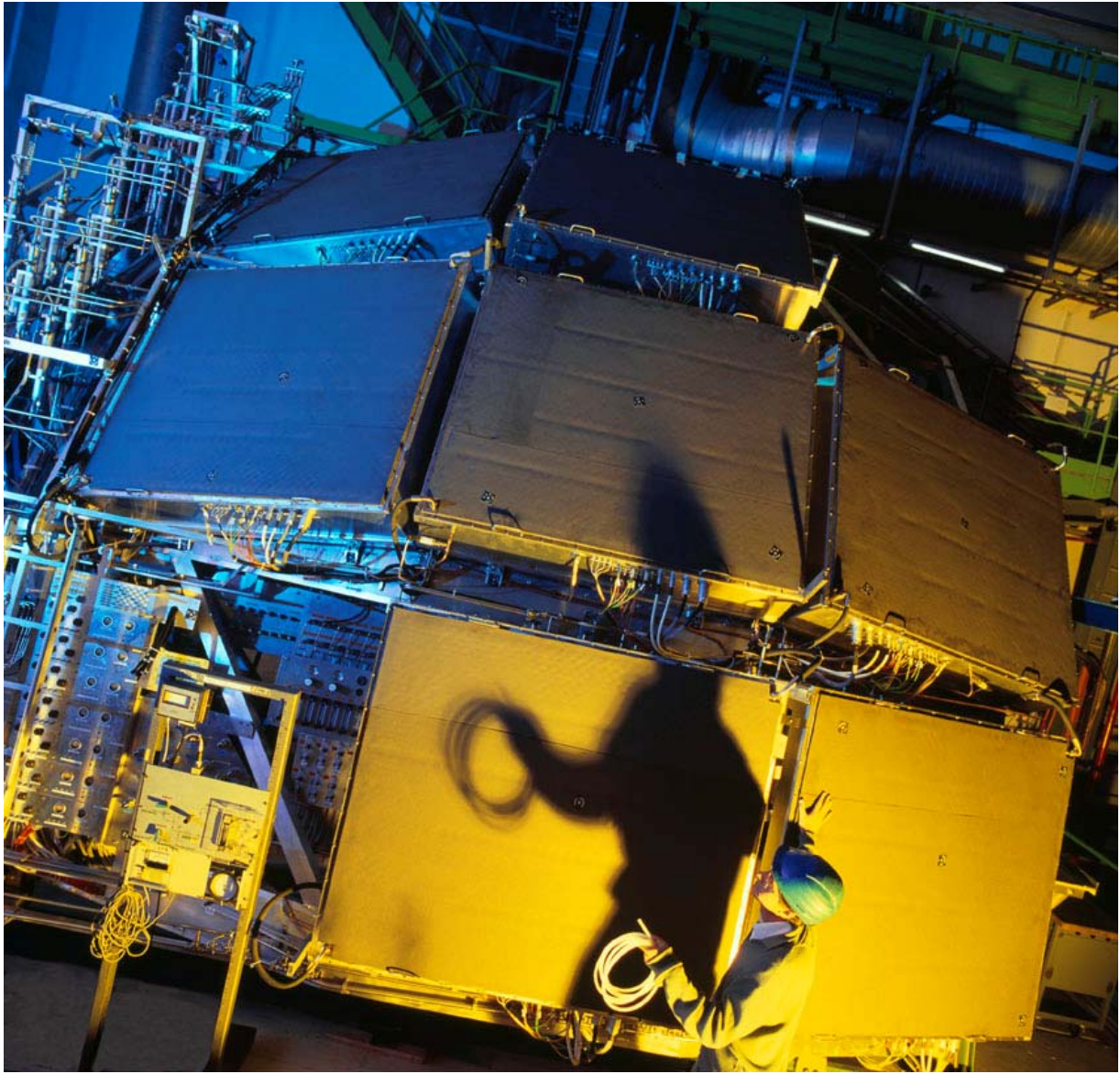

Fig 9: The HMPID prior to installation in the experiment (© A. Saba).

The Transition Radiation Detector (TRD)

Surrounding the TPC are 18 modules of a novel Transition Radiation Detector (TRD) which has the capability to distinguish between ultra-relativistic particles, such as the very 'special' electrons, and more conventional ones, e.g. pions. This detector has to perform on collisions with up to ~20 000 particles; furthermore, ALICE requires to have a fast(< 6μs) selection capability of electron and pion candidates, a world-novelty for this type of detector;
The high multiplicity requirement leads to a highly granular detector with detection elements ('pads') of 6cm$^2$ on average. The R&D phase addressed low-mass construction methods and understanding of the complex detector physics. The second area of development concerned the advanced readout electronics, most of which had to be integrated directly on the back of each of the 540 detector layers. This detector comprises a staggering 1.2 million channels spread over 700 m$^2$, by far the largest TRD ever constructed and possibly one of the most complex LHC detector System (Fig. 10).

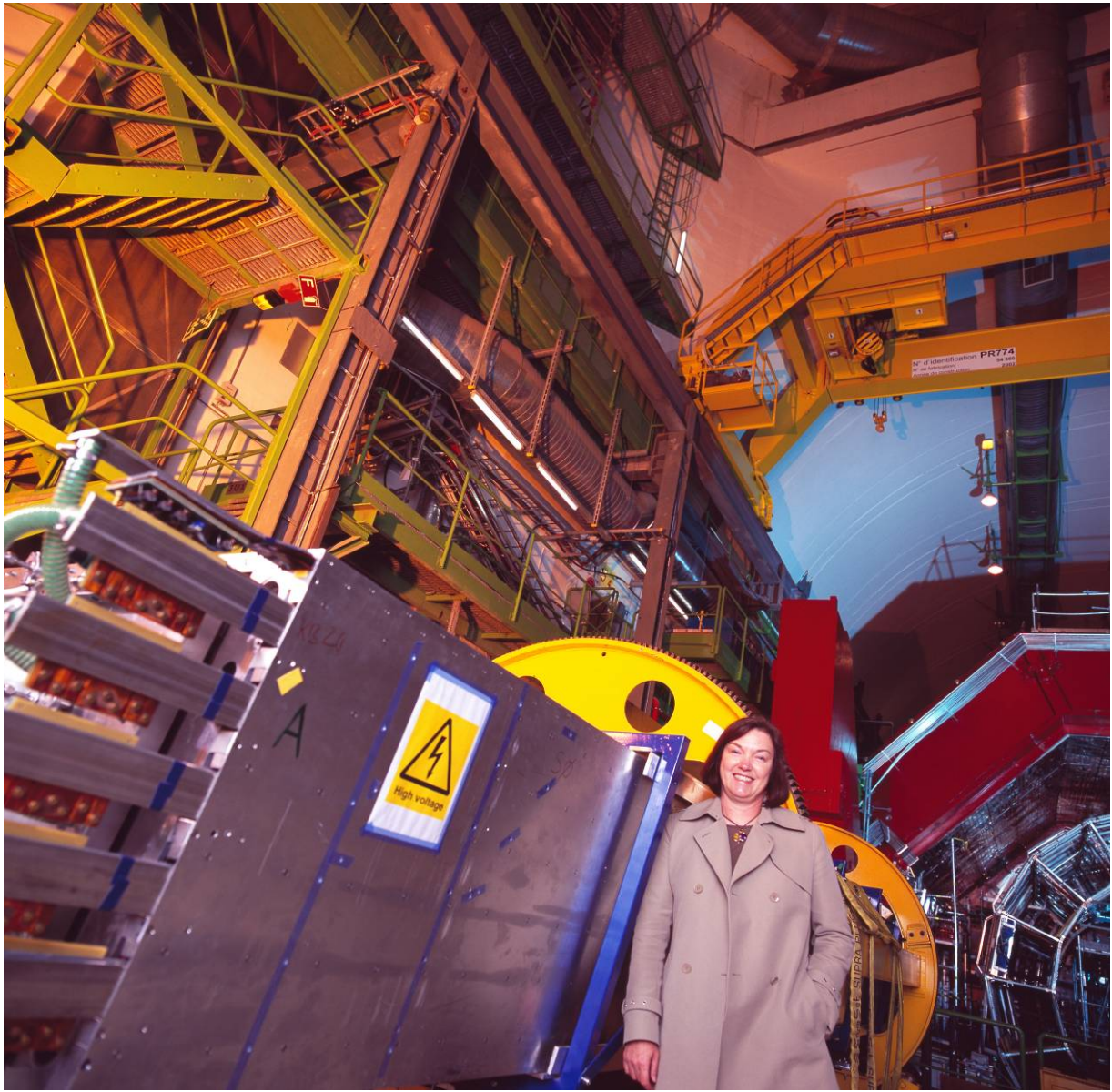

Fig 10: View of one of 18 TRD Detector units during installation (© CERN).

**Calorimeters**

Photons, spanning the range from thermal raditaion to hard QCD processes, as well as neutral mesons are measured in the small single-arm, high-resolution and high-granularity PHOS electromagnetic calorimeter. PHOS, which literally will 'take the temperature' of the collision, is located far from the vertex (4.6 m) and is made of dense scintillating crystals ($PbWO_4$) in order to cope with the large particle density. While otherwise very similar in design to the CMS crystal calorimeter described elsewhere in this book, it is cooled to -25$^0$C during operation to generate more light per incident energy and therefore improve the energy resolution. A set of multiwire chambers in front of PHOS helps in separating charged particles from photons (CPV).

High-energy partons kicked out by hard collisions will 'plow' through the primordial matter, loosing energy along the way before fragmenting into a spray of particles collectively called a 'jet'; these modified jets therefore probe the density and composition of the hot reaction zone. In order to enhance the capabilities for measuring jet properties, a second electromagnetic calorimeter (EMCal) will be installed in ALICE starting in 2008. The EMCal is a Pb-scintillator sampling calorimeter with longitudinal wavelength-shifting fibres, read out via avalanche photo diodes. Much larger than PHOS, but with lower granularity and energy resolution, it is optimized to measure jet production rates and jet characteristics in conjunction with the charged particle tracking in the other barrel detectors.

**Forward and trigger detectors**

A number of small and specialized detector systems are used for event selection or to measure global features of the reactions. Some of these detectors are shown in Fig. 11 and their location in ALICE can be traced in Fig. 1 with the help of the acronyms provided below:
- The collision time is measured with extreme precision ($< 2 \times 10^{-11}$ s) by two sets of 12 *Cherenkov counters* (fine mesh photomultipliers with fused quartz radiator) mounted tightly around the beam pipe (T0).
- Two arrays of segmented scintillator counters (V0) are used to select interactions and to reject beam related background events.
- An array of 60 large scintillators (ACORDE) on top of the L3 magnet will trigger on cosmic rays for calibration and alignment purposes, as well as for cosmic ray physics.
- The Forward Multiplicity Detector (FMD) provides information about the number and distribution of charged particles emerging from the reaction over an extended region, down to very small angles. These particles are counted in rings of silicon strip detectors located at three different positions along the beam pipe.
- The Photon Multiplicity Detector (PMD) measures the multiplicity and spatial distribution of photons in each single heavy ion collision. It consists of two planes of gas proportional counters with cellular honeycomb structure, preceded by two lead plates to convert the photons into electron pairs.
- Two sets of small, very compact calorimeters (ZDC) are located far inside the LHC machine tunnel (> 100 m) and very close to the beam direction to record neutral particles which emerge from heavy ion collisions in the forward direction.

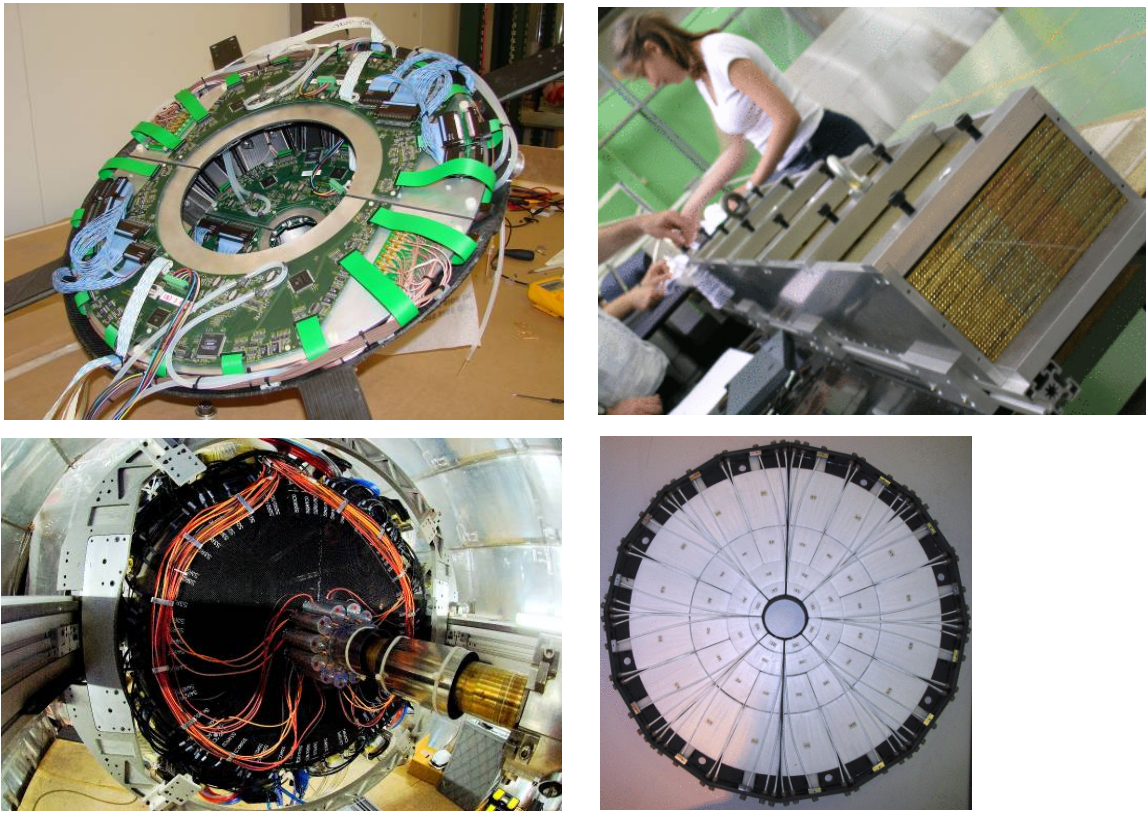

Fig 11: Forward and trigger detectors, clockwise from top left: two planes of the silicon FMD detector, one neutron ZDC, V0 scintillator array, start time T0 photomultipliers (© CERN).

**Muon Spectrometer**

On one side behind the central solenoid and at small angles between $2^0$ and $9^0$ relative to the beam direction is a muon spectrometer which includes a dipole magnet (4 MW power) generating a maximum field of 0.7 T. Several passive absorber systems (a hadron absorber close to the interaction point, a lead-steel-tungsten shield around the beam pipe and an iron wall) shield the spectrometer from most of the reaction products. The penetrating muons are measured in 10 planes of *cathode pad tracking chambers*, located between 5 m and 14 m from the interaction, with a precision of better than 100 µm. Again the relative low momentum of the muons of interest and the high particle density are the main challenges; therefore each chamber has two cathode planes, which are both read out to provide two-dimensional space information, and the chambers are made extremely thin and without metallic frames. The individual cathode pads range in size from 25 mm$^2$ close to the beam up to 5 cm$^2$ further away and cover 100 m$^2$ of active area with over 1 million active channels. Four trigger chambers are located at the end of the spectrometer, behind a 300 ton iron wall to select and trigger on pairs of muons from the decay of heavy quark particles. The chambers are made in the 'resistive plate' technology widely used by LHC experiments, and of modest granularity (20,000 channels covering 140 m$^2$). Figure 12 shows a 'fish eye' view of parts of the muon spectrometer in early 2008.

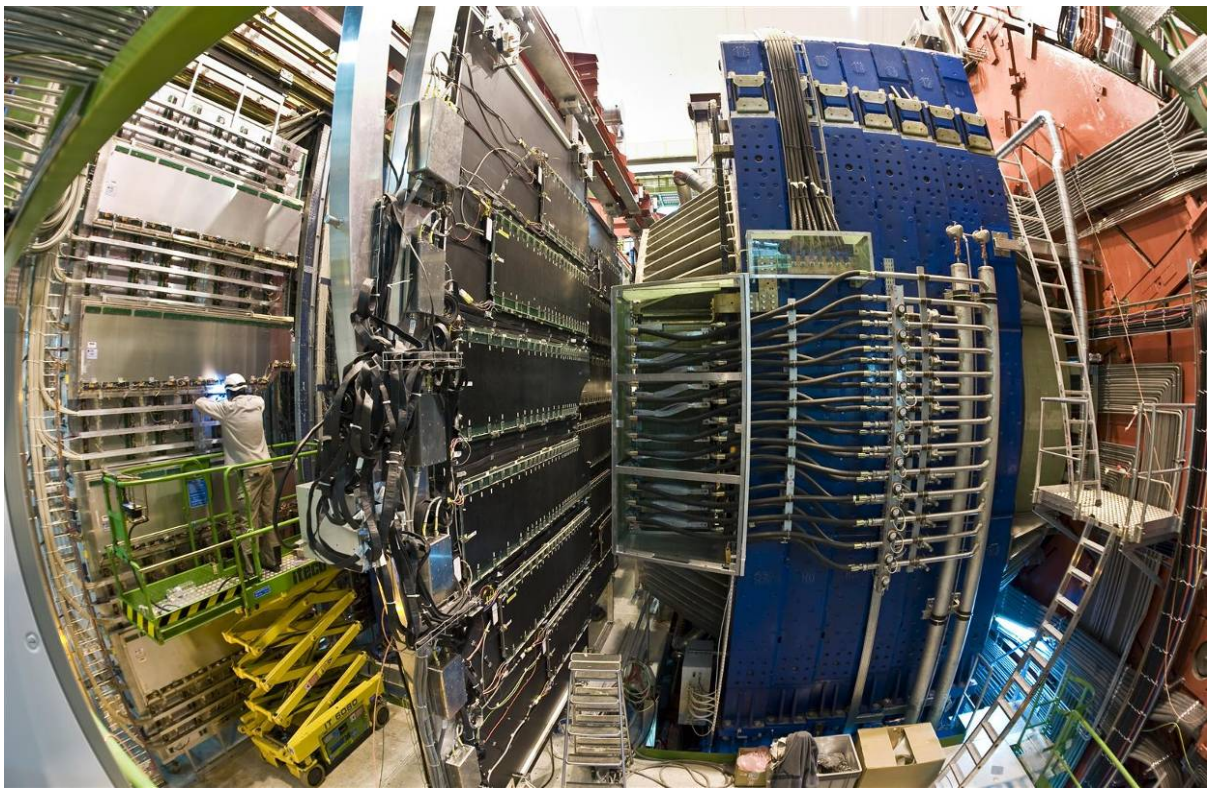

Fig 12: 'Fish eye' view of the forward muon spectrometer; visible are (from right to left) the main solenoid (red) and the muon dipole (blue) magnets, one of the tracking chamber planes and one of the triggering stations (©CERN).

In order to select the few muons from the thousands of other particles produced in each collision, an absorber is placed in front of the muon spectrometer. As the name implies, it should very effectively absorb hardons, while leaving muons through without much scattering (the usual task of such a 'hadron absorber'), but also at the same time minimize the sideways flux of shower debris (photons and neutrons in particular) which would harm the TPC performance. This seemingly innocuous object, seen during installation in Fig. 13, is in reality the most complex and finely tuned particle absorber ever built: it weighs 40 tons, has a length exceeding 4 meters and protrudes into the TPC closely to the collision point. A composite structure was developed, consisting of an inner tungsten core surrounding the beam pipe, followed by a conical high-density carbon absorber in the acceptance of the spectrometer to minimize multiple scattering of muons, followed by a lead-layer absorbing photons and an outer mantel of boron-loaded polyethylene to absorb neutrons. The hole assembly was literally 'cast in concrete' by pouring in mortar to fill out less critical spaces. From an engineering point of view a performance-cost optimization was a key issue, paying attention to submillimeter tolerances, fabrication and assembly procedures.

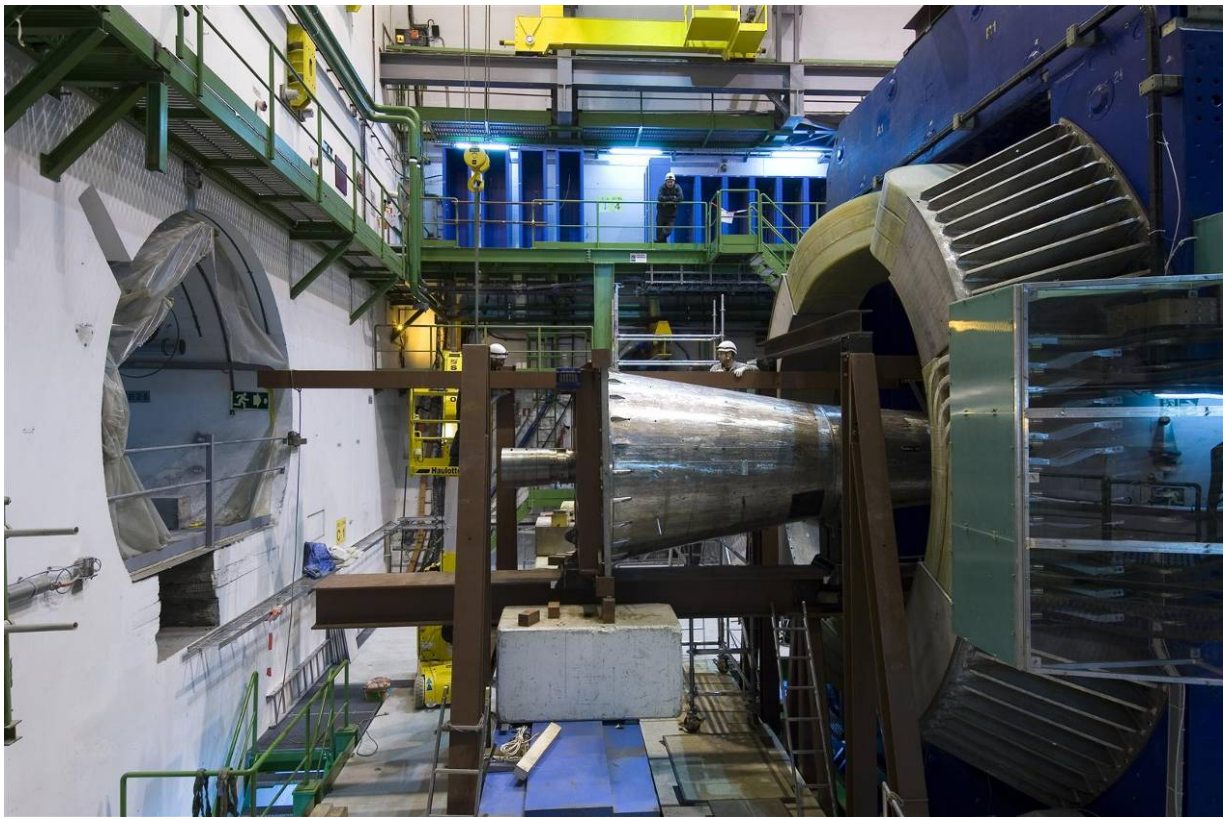

Fig 13: The conical hadron absorber slowly being pushed through the magnet of the muon spectrometer towards its final position (© CERN).

**Fitting the pieces together: engineering and detector integration**

Designing and constructing novel particle detectors was only part of the problem; building the mechanical supports and tools, fitting in the services and cables and, finally, assembling the pieces together was another, initially much underestimated challenge.
One example of sophisticated mechanical engineering in ALICE is the 'Space Frame', a 9-m-diameter tubular stainless-steel structure that houses the 80 tons of ITS, TPC, TRD, TOF and HMPID (Fig. 14). The collaboration insisted on a 'mass-less' structure maximizing active detector area and minimizing particle showers created in the support. During six months of optimization its weight was reduced by a factor 2, while at the same time improved understanding of the detectors raised their weight by 50%! The resulting, complex, welded tubular structure deforms by up to 12 mm when loaded. Laser and CCD camera angular monitors (36 in total) were specially developed and installed to provide a permanent monitor of the deflection and the feedback for adjustments.

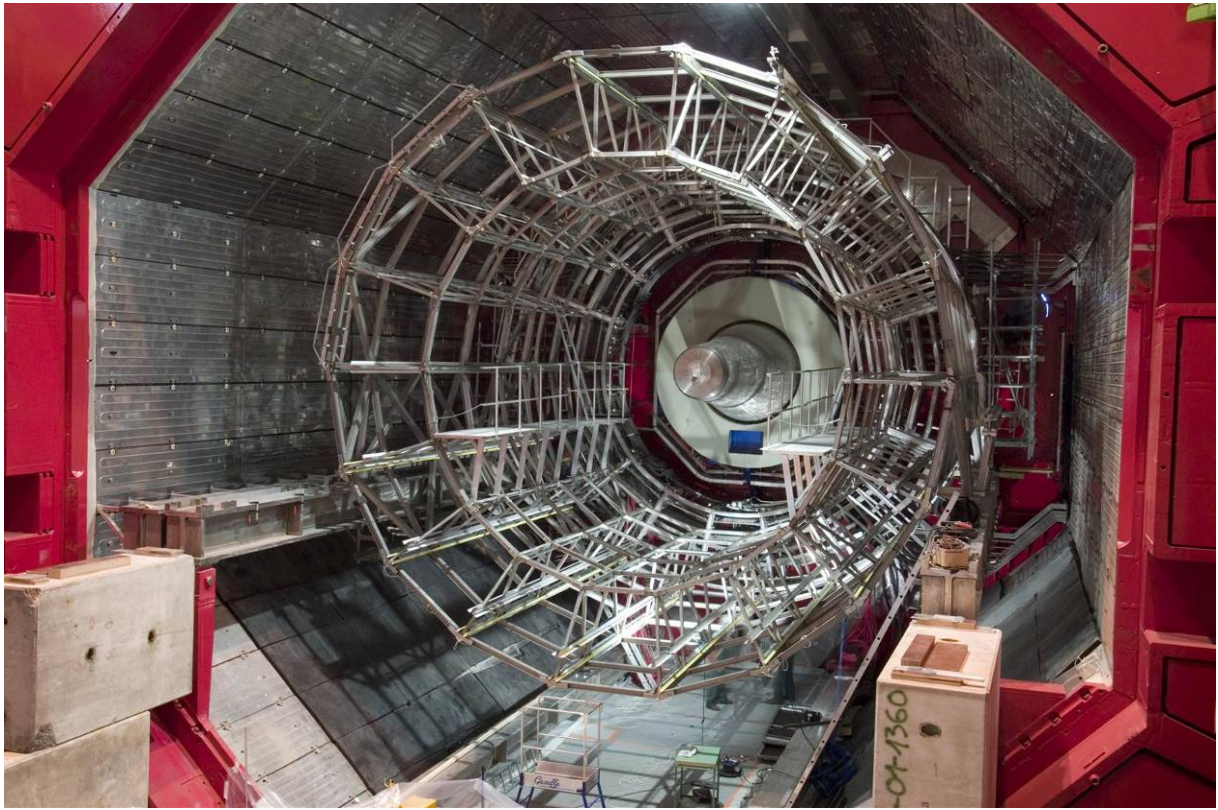

Fig 14: Spaceframe during installation. This stainless steel skeleton supports 80 tons of detectors, stressing its trusses close to the elastic limit. The mechanical integrity was load-tested on the surface prior to final installation (©CERN).

Two features of ALICE, not found in the other LHC detectors, are at the origin of several unusual installation challenges:

- the asymmetric layout with the large magnetic Muon Spectrometer constraints the installation of and access to the central detectors: the 200 tons of 'central' detectors can only be placed from the opposite side;
- these central detectors placed inside the L3 magnet can only be supported from the mechanically rigid iron crowns at the ends of the magnet, separated by a distance of 15 meters.

These seemingly innocent issues, discussed below, kept several brilliant and creative engineers occupied for the better part of five years.

The asymmetry imposed by the Muon spectrometer on the overall design was not accepted lightly. The alternative implied however displacing the 900-ton muon dipole magnet, the 300 ton muon filter, the muon detectors together with a large section of the delicate Be-vacuum chamber. The final verdict was rather clear: the complex, delicate one-sided installation represented the lesser evil.

First, the Muon Spectrometer was installed in its final, fixed position together with the hadron absorber. A 'ballet' of ITS and TPC motions had to be minutely orchestrated to allow the installation of the vacuum chamber, the independent Pixel and ITS detectors and, finally, the TPC. Connecting the detectors to cables, gas and cooling lines required to place these detectors at various intermediate positions. This was not only complex and delicate, it was potentially dangerous: the movement of the detectors caused large, up to 5 mm vertical deformations of the supports, while the vacuum chamber, attached to the detectors, was limited to excursions of less than 2.5. mm. This installation scenario could literally 'make or break' the ALICE experiment; it was reviewed by many committees, dress-rehearsed on the surface with many of the final components, monitored with strain gauges, feeler gauges, cross-checked by survey teams, engineers and physicists… It took the better part of nine months during 2007 to install these systems. Using a variety of tools from the inevitable duct tape to a dentist's drill for final dimensional adjustments, the operation was completed successful and in time as can be seen in Fig 15, which shows the experiment shortly before completion in early 2008, essentially ready to accept beams from the LHC.

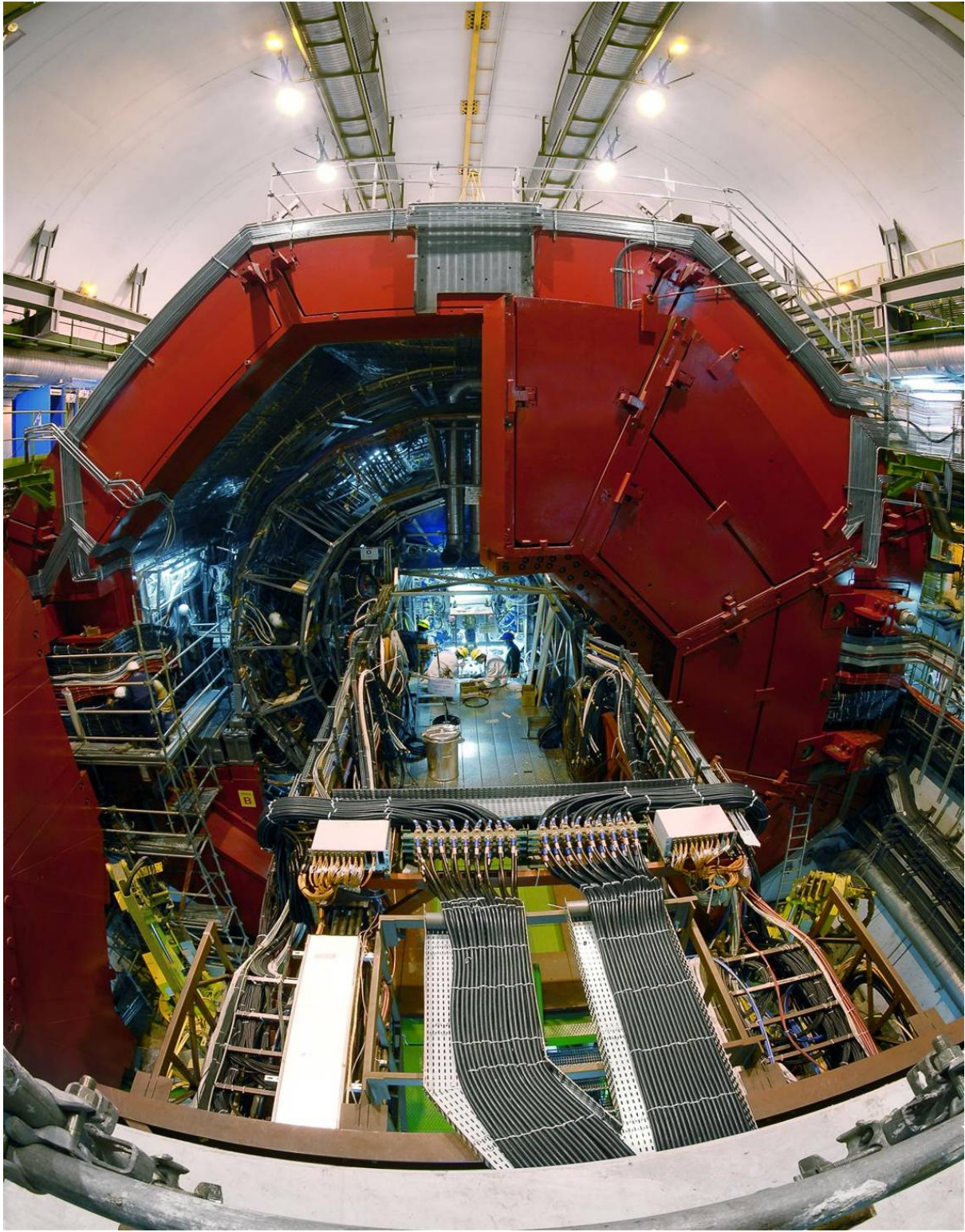

Fig 15: Front view of the L3 magnet, with its doors partially open. A bridge guides services, power cables, fluids and gases into the central detector. The silicon tracker is no longer visible and even the large TPC is mostly obscured. The stainless steel space-frame structure which supports most of the central detectors is partially filled with TOF and TRD modules(©CERN).

**Looking forward**

New technology, skilful engineering, and critical design decisions have led to a state-of-the-art detector that will be up to the task of observing the primordial matter created by heavy ion collisions in the LHC.  ALICE is the first truly universal 'general purpose' heavy ion experiment, which combines in a single detector most of the capabilities assigned in the past to several more specialised experiments. Incorporating the fruits of many years of R&D effort dedicated specifically to meeting the numerous challenges posed by the physics of nuclear collisions at the LHC, it is ready and well prepared, after more than 15 years of design and construction, to explore the 'little bang' and enter ALICE's wonderland of physics at the LHC.